\documentclass[letterpaper, 10 pt, conference]{ieeeconf}  

\IEEEoverridecommandlockouts                              
\usepackage{mathrsfs}
\usepackage{graphics} 
\usepackage{epsfig} 
\usepackage{amsmath} 
\usepackage{amssymb}  

\usepackage{url}
\usepackage{verbatim} 
\usepackage{soul, color}
\usepackage{lmodern}
\usepackage{fancyhdr}
\usepackage[utf8]{inputenc}
\usepackage{fourier} 
\usepackage{array}
\usepackage{makecell}

\usepackage{booktabs}
\usepackage{subfig}
\usepackage{lipsum}

\usepackage{algorithmicx}
\usepackage[ruled]{algorithm}
\usepackage[noend]{algpseudocode}
\usepackage{lineno}

\algnewcommand{\LineComment}[1]{\State \(\triangleright\) #1}

\makeatletter

\newcommand{\Rom}[1]{\expandafter\@slowromancap\romannumeral #1@}
\makeatother

\title{\LARGE \bf
ExplorerTree: a focus+context exploration approach for 2D embeddings*
}

\author{ \parbox{3 in}{\centering Wilson E. Marcílio-Jr, Danilo M. Eler
        \thanks{*This is the preprint version of a published work. DOI: https://doi.org/10.1016/j.bdr.2021.100239}\\
        Dept. of Mathematics and Computer Science\\
        São Paulo State University \\
        Presidente Prudente, SP, Brazil \\
        {\tt\small\{wilson.marcilio, danilo.eler\}@unesp.br}}

        \parbox{3 in}{ \centering Fernando V. Paulovich\\
        Faculty of Computer Science\\
        Dalhousie University\\
        Halifax, NS, Canada\\
        {\tt\small paulovich@dal.ca}}\\ \\
        
        \parbox{3 in}{ \centering José F. Rodrigues-Jr\\
        Inst. of Mathematics and Computer Sciences\\
        University of São Paulo\\
        São Carlos, SP, Brazil\\
        {\tt\small junio@icmc.usp.br}}
        
        \parbox{3 in}{\centering Almir O. Artero\\
        Dept. of Mathematics and Computer Science\\
        São Paulo State University \\
        Presidente Prudente, SP, Brazil \\
        {\tt\small almir.artero@unesp.br}}
}

\begin{document}

\maketitle
\thispagestyle{plain}
\pagestyle{plain}

\begin{abstract}

In exploratory tasks involving high-dimensional datasets, dimensionality reduction (DR) techniques help analysts to discover patterns and other useful information. Although scatter plot representations of DR results allow for cluster identification and similarity analysis, such a visual metaphor presents problems when the number of instances of the dataset increases, resulting in cluttered visualizations. In this work, we propose a scatter plot-based multilevel approach to display DR results and address clutter-related problems when visualizing large datasets, together with the definition of a methodology to use focus+context interaction on non-hierarchical embeddings. The proposed technique, called ExplorerTree, uses a sampling selection technique on scatter plots to reduce visual clutter and guide users through exploratory tasks. We demonstrate ExplorerTree's effectiveness through a use case, where we visually explore activation images of the convolutional layers of a neural network. Finally, we also conducted a user experiment to evaluate ExplorerTree's ability to convey embedding structures using different sampling strategies.\\

\end{abstract}

\begin{keywords}

Dimensionality Reduction, Visualization, Scatter-plot, Focus+context

\end{keywords}

\section{Introduction}
\label{sec:introduction}

The increasing amount of high-dimensional data has led to the development of techniques that try to extract patterns present in these datasets. For example, Multidimensional Projection (MP) or Dimensionality Reduction (DR)~\cite{Maaten_2008,Paulovich_2008,Joia_2011,McInnes2018} techniques produce low-dimensional representations (or embeddings) that respect similarity relationships among data points~\cite{Paulovich_2010}. The results of these techniques are usually visualized as scatter plots, where spatial proximity depicts similarity. For small datasets, scatter plots can rapidly convey clusters and patterns present in the data. However, the analysis of big datasets tends to cause problems due to the non-visual scalability of scatter plot representations. Accordingly, interaction mechanisms are applied to ease data analysis, such as hierarchical exploration. These hierarchical approaches have the advantage of providing information on demand, according to information demand. For example, in techniques such as \textit{HiPP}~\cite{Paulovich_Minghim_2008} and \textit{h-SNE}~\cite{Hollt2019}, users visualize dominant structures, then successive refinements on the visual layouts make data observations accessible.

Interaction mechanisms, such as brushing and linking, coordination, and focus+context, play an essential role in providing more straightforward exploration and helping users cope with cognitive overload. The focus+context concept~\cite{Munzner2015}, for example, can be naturally used in hierarchical representations of DR results since the most fine-grained information corresponds to the focus, and the information that is not necessary at the moment corresponds to the context. Höllt et al.~\cite{Hollt2019} proposed the focus+context concept for hierarchical embeddings (results generated by hierarchical dimensionality reduction techniques) by defining a set of requirements to support exploration. However, their work does not address DR techniques that operate in one single level of detail. Besides that,  there is no study combining focus+context exploration in any embedding results with concerns about visual scalability.

In this work, we propose the ExplorerTree, an approach to hierarchically explore dimensionality reduction results using the focus+context exploration approach that accounts for concerns of visual scalability of scatter plot-based representations. Throughout the exploratory process, users receive visual cues and important structures about the projection. Further, our approach extends the hierarchical exploration concept proposed by Höllt et al.~\cite{Hollt2019} even to DR techniques that cannot be extrapolated to hierarchical versions~\cite{NonatoAupetit}. Since the embedding's hierarchical structure is created based on a sampling selection mechanism~\cite{MarcilioJr2020} that preserves the scatter plots visualization overview (e.g., class boundaries and outliers), users visualize the main structures with reduced cognitive effort induced by overlapping visual markers. The sampled data points guide users throughout the exploratory process using the overview-first \& details-on-demand mantra~\cite{Shneiderman_1996}. More data points are projected onto the visual space to provide finer-grained information based on the interaction with the representative data points, that address the scatter plot metaphor's well-known difficulties related to visual clutter. 

Throughout the paper, we propose new requirements for hierarchical exploration, accounting for visual scalability. Then, we demonstrate our encoding strategies to represent clusters' density, class variability, and content information for each node of the hierarchy generated from the successive sampling selection. Finally, we provide a use case to demonstrate our approach's viability by exploring the activation images of convolutional layers of a convolutional neural network.

Summarily, the main contributions of this work are:

\begin{enumerate}
\item A hierarchical exploration approach of any embedding using focus+context concept, accounting for visual scalability; 
\item A design mechanism to reduce visual clutter on hierarchical representations of dimensionality reduction results;
\item Requirements to facilitate exploration of 2D embeddings considering visual scalability.
\end{enumerate}

This paper is structured as follows. Section~\ref{sec:related-works} presents the related works. In Section~\ref{sec:requirements}, we discuss the requirements for the hierarchical exploration of multidimensional embeddings. In Section~\ref{sec:explorertree}, we explain the ExplorerTree hierarchical construction mechanism and visualization design, and how we choose the sampling selection mechanism. Section~\ref{sec:use-case} provides a use case using ExplorerTree to explore filter activations of a convolutional neural network. In Section~\ref{sec:discussion}, we discuss the technique and its limitations. Finally, we conclude our work in Section~\ref{sec:conclusion}.

\section{Related Works}
\label{sec:related-works}

Many dimensionality reduction techniques have been proposed through the years, presenting advantages in various aspects. For example, LSP~\cite{Paulovich_2008} presents promising results for preserving neighborhood relations, while t-SNE~\cite{Maaten_2008} can efficiently separate classes on the projected space. Espadoto et al.~\cite{Espadoto2019} provide a useful categorization of dimensionality reduction techniques. From the visualization point of view, Nonato and Aupetit~\cite{NonatoAupetit} present an excellent categorization of multidimensional projection techniques, linking visualization with distortion, tasks, and many other aspects. Although applied in a variety of problems, dimensionality reduction techniques generate cluttered layouts for big datasets, which difficult analysis and require high cognitive overload.

Hierarchical and interaction mechanisms facilitate exploratory tasks and knowledge discovery, besides minimizing cluttered layouts when dealing with large datasets. For example, \textit{InfoSky}~\cite{Andrews_2002} hierarchically explores document collections by mapping documents to star \textit{glyphs} and using polygons to depict similar documents and provide separation among clusters. Users change the level of detail according to demand following a telescope metaphor. \textit{MDSteer}~\cite{Williams2004} computes an MDS embedding, showing promising results in the first hierarchical levels. However, overplotting is still present when reaching the lowest levels of the tree structure. Besides that, there is no additional information to help users in the exploration tasks. \textit{HIPP}~\cite{Paulovich_Minghim_2008} uses representatives extracted from a clustering process to generate a hierarchical layout. The technique uses LSP for positioning data instances and rearranges the layout on the first hierarchical level to reduce overlap among nodes (circles with varying sizes that encode clusters). Although this process reduces the overplotting by reducing overlap, it modifies the global relations among data samples. For instance, meaningful hierarchical exploration strategies are one of the challenges of single-cell RNA sequencing data~\cite{Lahnemann2020}, in which global relations play an important role. Pezzotti et al.~\cite{Pezzotti_2016} proposed a technique called \textit{h-SNE}, which uses \textit{t-SNE}~\cite{Maaten_2008} to perform projection and builds the hierarchy based on random walks on a neighborhood graph. Such characteristic allows exploration according to information demand. However, the generated layout can still be cluttered and confusing if these structures have a large size. Another related work, called Visual Super Tree~\cite{Silva2019}, employs focus+context exploration to reduce the burden when analyzing big datasets using the similarity tree metaphor. It employs node expansion and contraction to refine and hide information throughout the exploratory process. Besides being dependent on the number of clusters chosen to create the hierarchy (as in the HiPP technique), during the exploratory process, visual scalability can be a problem when dealing with big datasets that require successive node expansions. There is no proper way to differentiate structures in focus from structures in context during successive node expansions. While visualizing the expanded nodes in new windows addresses this problem, such an approach loses the context between data instances.

The focus+context concept for hierarchical multidimensional embeddings was defined by Höllt et al.~\cite{Hollt2019} through a set of design requirements. However, while the requirements (detailed in the next section) are useful for interaction/navigation aspects, there is still no discussion about the hierarchical exploration of multidimensional embeddings' visual scalability.

In this work, we address the problem of focus+context interaction in multidimensional embeddings and give attention to the visual scalability aspects during exploration. We propose three new requirements besides showing mechanisms to apply focus+context exploration to any multidimensional embedding, not only a few existing hierarchical ones. Our work also accounts for hierarchical dimensionality reduction techniques, making our approach applicable to all DR results represented by scatter plots.

\section{Requirements for focus+context exploration of hierarchical embeddings}
\label{sec:requirements}

A hierarchical exploration approach must follow a few design requirements to facilitate user analysis. In addition to the interaction (\textbf{I1} - \textbf{I9}) and visual (\textbf{V1} - \textbf{V4}) requirements proposed by Höllt et al.~\cite{Hollt2019} to support a compelling exploration of hierarchical embeddings using the focus+context concept (see Table~\ref{tab:Höllt-requirements}), in this work, we propose three new requirements (\textbf{R1} - \textbf{R3}) to deal with visual clutter in 2D embeddings visualized as scatter plots and to guide users through exploratory processes.

\begin{table}[h!]
 \caption{Requirements for supporting the exploration of hierarchical embeddings using Focus+Context concept. Interaction (\textbf{I1}-\textbf{I9}) and Visual (\textbf{V1}-\textbf{V4}) were proposed by T. Höllt et al.~\cite{Hollt2019}.}
 \label{tab:Höllt-requirements}
 \scriptsize
 \begin{center}
   \begin{tabular}{l|l}
   \toprule
     Requirement & Description \\
   \midrule
   	\textbf{I1} & request more detail for all data\\
   	\textbf{I2} & request less detail for all data\\
   	\textbf{I3} & return to the initial state\\
   	\textbf{I4} & define an area of interest (focus)\\
   	\textbf{I5a} & change the focus to a subset of the current focus\\
   	\textbf{I5b} & change the focus to a different set of points\\
   	\textbf{I6} & request more detail for the focus\\
   	\textbf{I7} & request less detail for the focus\\
   	\textbf{I8} & create a second focus for comparison\\
   	\textbf{I9} & resolve the second focus\\   
   \midrule
    \textbf{V1} & the focus must use extended space in the visualization\\
    \textbf{V2} & the focus must be separated from the context\\
    \textbf{V3} & maintain connections between focus and context \\
    \textbf{V4} & different hierarchy levels must be identifiable\\
   \midrule
    \textbf{R1} & reduce visual disorder (clutter)\\
    \textbf{R2} & show information about the clusters\\
    \textbf{R3} & reduce overlaps among graphical markers\\
    \bottomrule
   \end{tabular}
 \end{center}

\end{table}

The requirements of this study deal with the visual scalability aspects faced when exploring embeddings represented by scatter plots. Specifically, requirements \textbf{R1} and \textbf{R3} deal with the visual representation's ability to communicate clusters and patterns in the data. While these two requirements may seem very similar at first, they acknowledge different problems usually present in traditional scatter plot representations. \textbf{R1} \textit{(reduce visual disorder (clutter))} considers the problem when a projection is too cluttered with visual markers, being difficult to identify patterns and clusters. On the other hand, \textbf{R3} \textit{(reduce overlaps among visual markers)} helps users explore clusters without losing track of the relationship imposed by the dimensionality reduction process. While readers may think that reducing overlapping the visual disorder will also be reduced, the literature~\cite{Marcilio2019} shows that the existing removal algorithms cannot reduce the overlapping of big datasets without imposing severe issues to the similarity and neighborhood structures of the projected dataset. Finally, \textbf{R2} (\textit{show information about the clusters}) helps users navigate the dataset using visual cues of the cluster's content. Such a requirement also helps to obtain some of the information lost during sampling selection.

In summary, a big part of our work is related to deal with visual scalability problems of scatter plot representations produced by dimensionality reduction methods. While removing all of the overlappings of a projection results in layouts with issues regarding similarity and neighborhood structures~\cite{Marcilio2019}, the visual exploration is usually performed by looking at dominant structures and refining the search according to the user's needs. That is, we argue that efforts to reduce overplotting globally can be meaningless when taking the natural exploration approach into account. In this work, we use sampling selection and overlapping removal approaches on a hierarchical representation of scatter plots; the aim is to provide a facilitated exploration of high-dimensional data, which comprises our design mechanism to reduce visual clutter on hierarchical representations of dimensionality reduction results.

\section{ExplorerTree}
\label{sec:explorertree}



\begin{figure}[!htb]
\centering
\includegraphics[width=\linewidth]{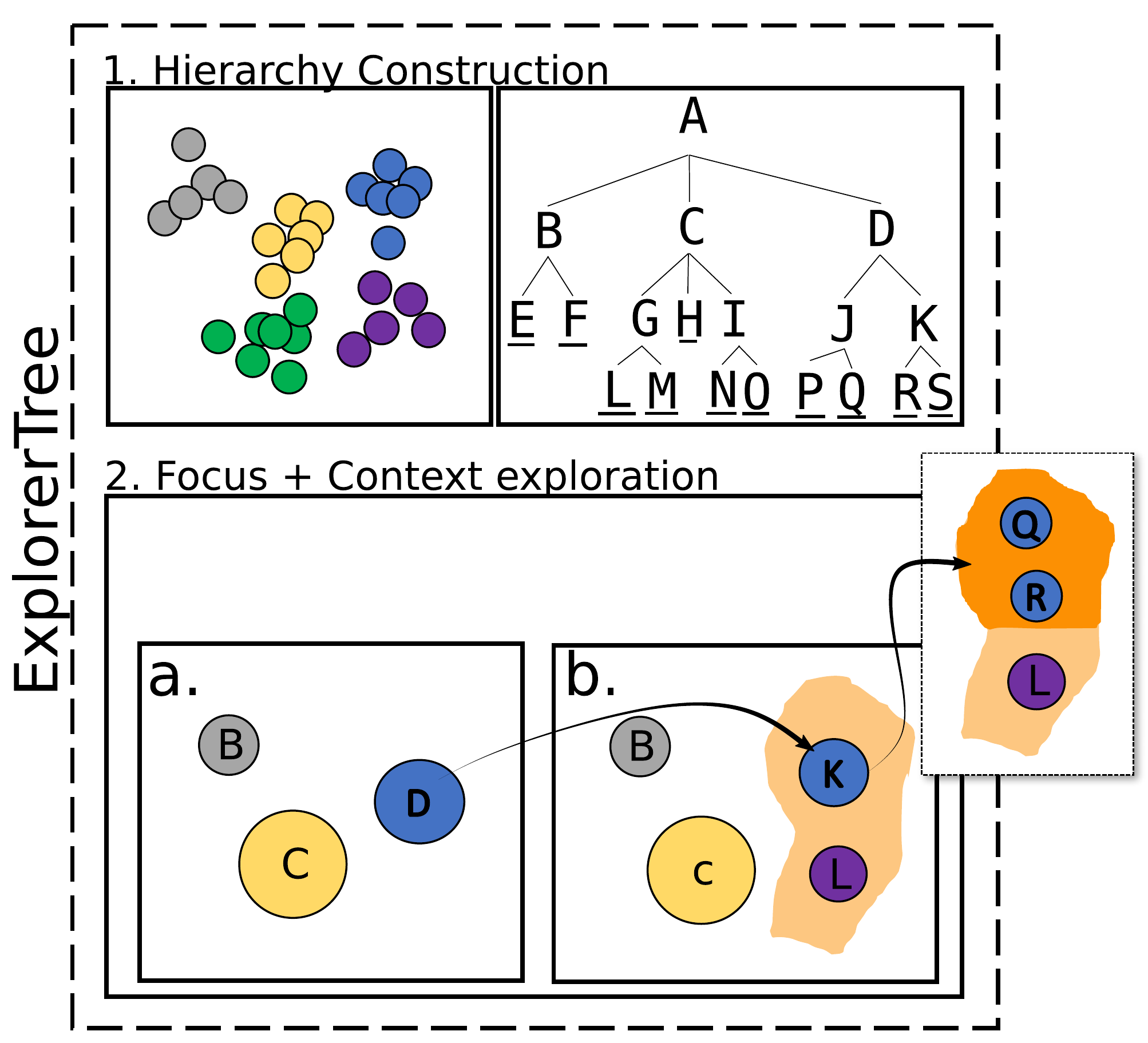}
\caption{Schematic view of ExplorerTree. Our approach takes points in $\mathbb{R}^2$ to create a tree structure based on representative selection \textbf{(1)}, then exploration is performed using the Focus+Context concept \textbf{(2)}. In the visual representation, a color scale encodes hierarchy levels while visual summaries are employed to understand clusters’ content.}\label{fig:scheme-approach}
\end{figure}

Using the design requirements, we define a hierarchical exploration approach through successive sampling and clustering procedures on the embedding space. Figure~\ref{fig:scheme-approach} shows representative selection (or sampling) from $\mathbb{R}^2$ embedding to create a hierarchical structure on the data \textbf{(1)}; then, the second step \textbf{(2)} deals with the use of visual variables (e.g., area and color), interaction mechanisms (e.g., overview-first \& details-on-demand and focus+context), visual summaries and overlap removal mechanisms to provide a facilitated exploration of the hierarchy.

A hierarchical structure for a 2D embedding is created based on sampling and clustering from the data points – we use the terms \textit{sampling} and \textit{representative selection} interchangeably. Using sampling, we reduce visual complexity – since fewer data points are projected onto the plane – while maintaining useful information about the embedding structures (requirement \textbf{R1}), that is, class boundaries and density information. The first sampling process results in representative data points that generate the nodes/clusters of the first hierarchical level. According to Euclidean distance, the non-representative data points are associated with the closest representative data point cluster in these clusters. Then, the nodes/clusters computed from the first sampling process result in the input of new sampling processes to induce a hierarchical structure on the scatter plot representation.

\begin{figure}[!htp]
\centering
\includegraphics[width=\linewidth]{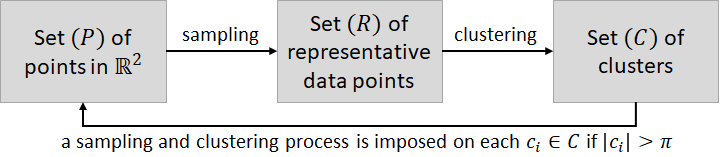}
\caption{Sequence of steps to create a tree structure on data points in $\mathbb{R}^2$. Representatives are sampled from a set of points and used to impose a partition on the dataset, generating a set of clusters. The process (selecting representatives and defining partition) is repeated for each cluster whose size is greater than a pre-specified parameter ($\pi$).}\label{fig:workflow}
\end{figure}

Figure~\ref{fig:workflow} illustrates the process of imposing a hierarchical structure on points in $\mathbb{R}^2$. At first, the set of points consists of the whole dataset. Then, the sampling and clustering algorithms return the clusters/nodes of the first hierarchy level. Such a process is repeated for each cluster returned in the last component until a cluster has fewer data points than a pre-defined parameter, represented by $\pi$.

We did not use hierarchical clustering for sampling selection because much information is lost when projecting only the medoids. Instead, we aim to preserve, as discussed in Section~\ref{sec:representative-selection}, the overall structure and the context of the scatter plot layout and ensure good representativity of the whole dataset.

\subsection{Representative selection}
\label{sec:representative-selection}

The representative selection is one of the essential steps of ExplorerTree. A suitable representative/sampling selection technique selects data points that cover the embedding with low redundancy. A more trustful exploration is achieved if a sampling selection technique preserves the structures and class boundaries of a scatter plot representation. 

\subsubsection{Visual quality using different sampling strategies}

A sampling technique must cover the structures presented in the visual space and do not present redundancy among the sampled data instances to guarantee good preservation of structures and not deceive users throughout the exploratory process. In the \textbf{Supplementary File (Figure 2)}, we numerically evaluated these aspects for SADIRE~\cite{MarcilioJr2020}, k-Means~\cite{Steinbach2000}, Reservoir~\cite{Vitter1985}, and CSM~\cite{Joia2015} techniques. The experiments show decent run-time executions for SADIRE (losing only to Reservoir) while surpassing all the other techniques regarding coverage and redundancy aspects.

\begin{figure}[htp]
\centering
\includegraphics[width=\linewidth]{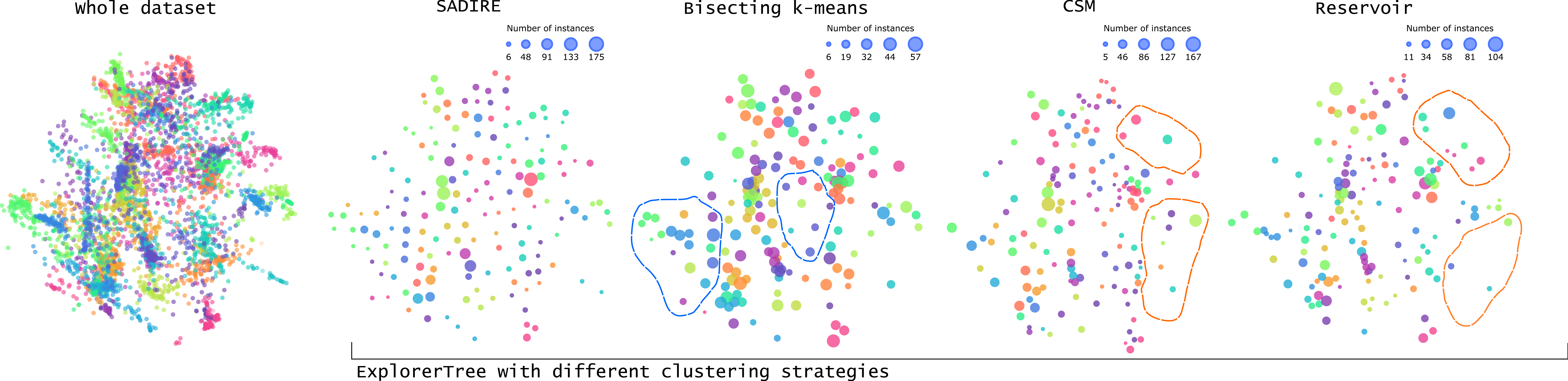}
\caption{ExplorerTree using different sampling algorithms. On the left, the traditional scatter-plot representation. On the right, the first level of ExplorerTree using four sampling strategies.}\label{fig:explorertree-different-algorithms}
\end{figure}

Figure~\ref{fig:explorertree-different-algorithms} shows an example of SADIRE's suitability against the other techniques. On the left, LSP~\cite{Paulovich_2008} projection of the Corel dataset with five thousand data instances colored according to their class. On the right, we show the first level of ExplorerTree (detailed explained in the next section) on the same dataset using the sampling strategies. Unlike the other techniques, SADIRE covers all the projected space with sampled data instances while the circle area correctly shows the densest parts of the projection. Reservoir and CSM lose many structures that SADIRE correctly encodes – see these structures highlighted in orange. Finally, Bisecting k-means does not encode well the density information, aggregating data instances that do not correspond to the projection's density – see the structures highlighted in blue. \textbf{Supplementary File's Figure 6} shows the same analysis for other datasets.

We performed a user experiment using these sampling strategies to verify how participants evaluate ExplorerTree first hierarchical level under different conditions. The details of this user experiment are in \textbf{Section 5} of \textbf{Supplementary File}. Nevertheless, the participants assigned higher scores for ExplorerTree using SADIRE and Bisecting k-Means, which justifies our choice for SADIRE due to its rapid run-time execution needed for interactive purposes.

\subsubsection{Selecting candidates for representative set using SADIRE}
\label{sec:select-candidates}

\begin{figure}[!htb]
\centering
\includegraphics[width=\linewidth]{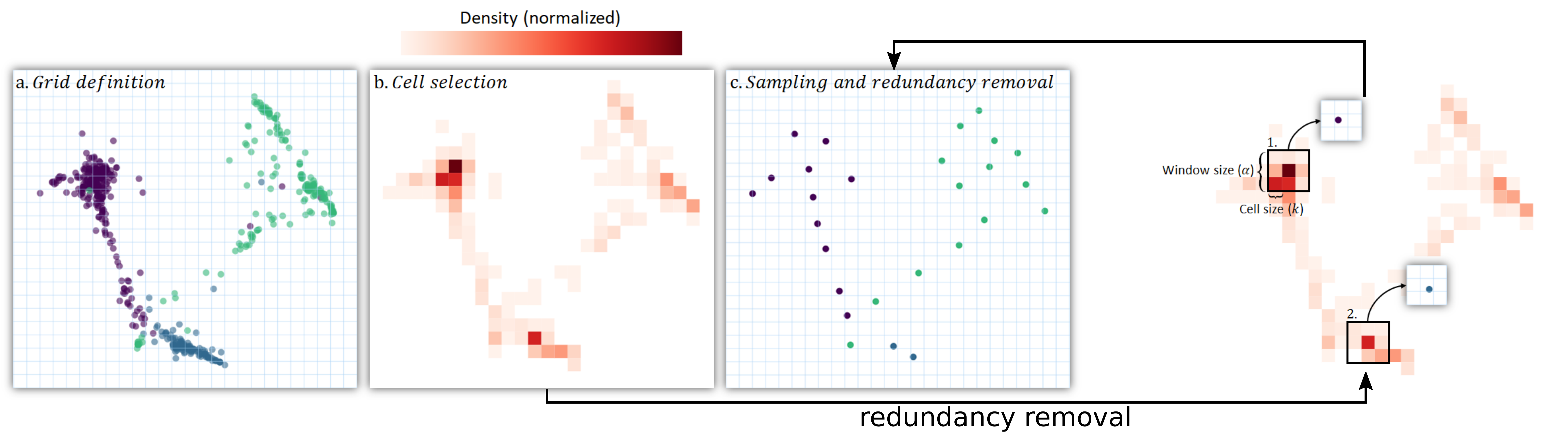}
\caption{SADIRE's main execution steps. (a) A grid of cells with size $k \times k$ divides the $\mathbb{R}^2$ space, the density of each cell is also stored; (b) Only cells with at least one point are retrieved; (c) For each cell, the medoid selected as a representative data point. Between steps (a) and (b), SADIRE applies a redundancy removal mechanism -- in this case, a window of size $\alpha = 3$ removes redundant data points..}
\label{fig:first-step}
\end{figure}

Given a scatter plot in $\mathbb{R}^2$, SADIRE uses a grid of cells with size $k\times k$ to divide the projected space and retrieves the medoid from each cell with at least one data point. Figure~\ref{fig:first-step} illustrates the main steps of SADIRE sampling technique. After the projection of data points onto $\mathbb{R}^2$, a grid with cell size $k\times k$ group too similar data points (a). Then, a set of candidates for representative data points consists of the cells with data points (b) – the density of each cell also has to be stored in this step, as depicted by color saturation. Finally, only the medoid of each cell is selected as the representative data point after redundancy removal (c).

\subsection{Hierarchy Construction}
\label{sec:tree-construction}

\begin{figure}[htp]
\centering
\includegraphics[width=\linewidth]{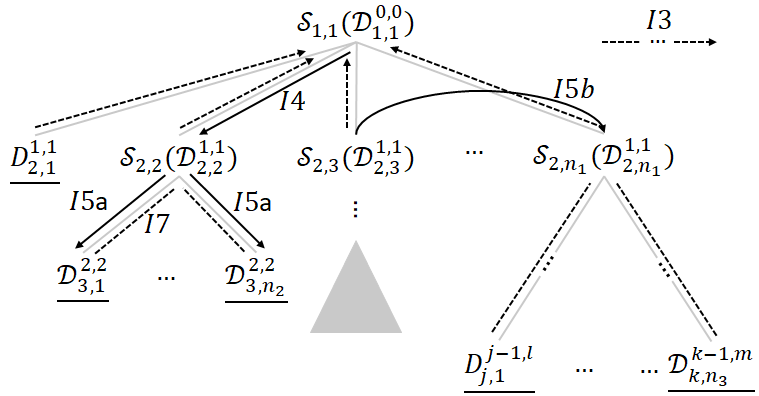}
\caption{Tree structure created by sampling on the cluster/nodes. The bars below nodes indicate leaf nodes, where ExplorerTree ends representative selection.}\label{fig:tree-structure}
\end{figure}

The sampling technique allows ExplorerTree to cluster data points and employ successive operations to create a hierarchical structure. Formally, let $\mathscr{D}^{0,0}_{1,1}$ be a set of points in $\mathbb{R}^2$ after dimensionality reduction with no imposed hierarchical structure, where $\mathscr{D}^{j-1,l}_{j,i}$ denotes a cluster $i$ in the hierarchy level $j$, and ($j-1, l$) denotes the level and the cluster index of its parent. The first sampling process ($\mathscr{S}_{1,1}$) imposes a partition on the cluster of the first level, i.e., the set of points in $\mathbb{R}^2$ after dimensionality reduction. Figure~\ref{fig:tree-structure} illustrates the sampling process, which returns $n$ ($2 \leq n \leq m$) clusters, where $m$ depends on the minimum size of data points allowed in each cluster, as detailed in Section~\ref{sec:considerations}. The sampling selection repeats for each cluster of the second level of the hierarchy -- for each $\mathscr{D}_{2,i}^{1,1}$. Notice that the index of any sampling process corresponds to the same cluster index, and the superscript indices of any cluster/node must match the subscript indexes of its parent. In the figure, gray edges indicate the parent-child relationship. Black arrows indicate the requirements, the bars below nodes indicate tree leaves, and dashed arrows indicate successive operations from the leaves to the root.

Generalizing, the sampling process ($S_{j,i}$) of an arbitrary cluster ($i$) in an arbitrary hierarchy level ($j$) extracts representative data points from the corresponding cluster on the corresponding level ($\mathscr{D}_{j,i}^{j-1,l}$). The result of sampling is a partition on the cluster, where the union of all generated subclusters results in the parent cluster ($\bigcup \mathscr{D}^{j, i}_{j+1, \lbrace K\rbrace} = \mathscr{D}_{j, i}^{j-1, l}$, where $\lbrace K\rbrace$ is the set of all subclusters), and two subclusters cannot share any data point ($\mathscr{D}_{j,a}\cap \mathscr{D}_{j,b} = \emptyset, \forall a\neq b$). 

Using the tree structure, we accomplish a few requirements of Table~\ref{tab:Höllt-requirements} and highlighted in Figure~\ref{fig:tree-structure}. For instance, successive resolving focus operations accomplish requirement \textbf{I3} (return to the initial state). Selection of representatives accomplishes requirement \textbf{I4} (define an area of interest). Requirements \textbf{I6} (request more detail for the focus) and \textbf{I7} (request less detail for the focus) are accomplished with interaction on the pre-defined clusters. Requirement \textbf{I5a} (change the focus to a subset of the current focus) is accomplished by interacting with the children of the focused node, i.e., by requesting focus on the focused node's children. Requirement \textbf{I5b} (change the focus to a different set of points) is accomplished by our technique when users decide to focus on two different clusters/nodes – a curved horizontal arrow indicates this. Requirement \textbf{I8} (create a second focus for comparison) is accomplished by only interacting with another node of interest. Finally, by resolving focus on the node selected for comparison, we accomplish requirement \textbf{I9}.

\subsubsection{Considerations}
\label{sec:considerations}

First, the number of sampling and clustering processes executed on the embedding depends on the minimum cluster size parameter ($\pi$), which influences in the number of hierarchy levels. That is, the bigger is $\pi$, the lower is the number of levels in the tree. Finally, requirement \textbf{R3} specifies that the overlap among markers must be decreased/removed. In this case, such an operation is performed based on user interaction when one reaches a leaf cluster. Overlap removal can be performed using various techniques – such as PRISM~\cite{Gansner2010}, VPSC~\cite{Dwyer2006}, ProjSnippet~\cite{GomezNieto2014}, and RWordle~\cite{Strobelt2012} – since leaf cluster have limited size and, consequently, there is no much concern about which one to apply~\cite{Marcilio2019}.  

During hierarchical exploration, SADIRE sampling selection mechanism is sufficient to reduce overlap among visual markers. However, when inspecting data observations on the lowest hierarchy levels, we cannot sample data points to facilitate exploration nor impose significant changes on the layout arrangement. However, it is interesting to know that overlapping removal on clusters with only a few data points does not impose significant changes in structural relations~\cite{Marcilio2019}. In this case, choosing a reasonable number of $\pi$ ($\sim 200$)  results in layout with reduced information loss after overlap removal. 

\begin{figure}[!htb]
    \centering
    \subfloat[Case 1: a partition generates two invalid cluster -- red and purple --, which are merged.]{\includegraphics[width=\linewidth]{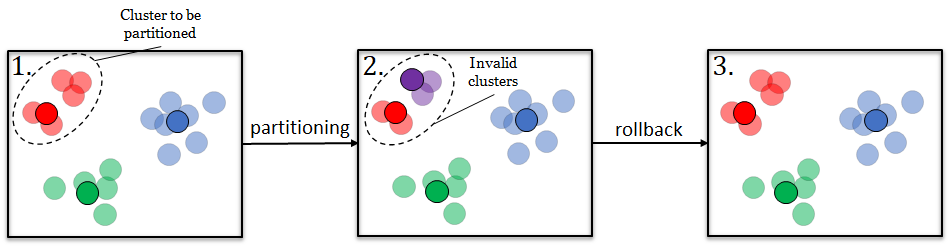}\label{fig:processo-fundir-caso1}}
    
    \subfloat[Case 2: a invalid cluster is merged to its nearest cluster.]{\includegraphics[width=\linewidth]{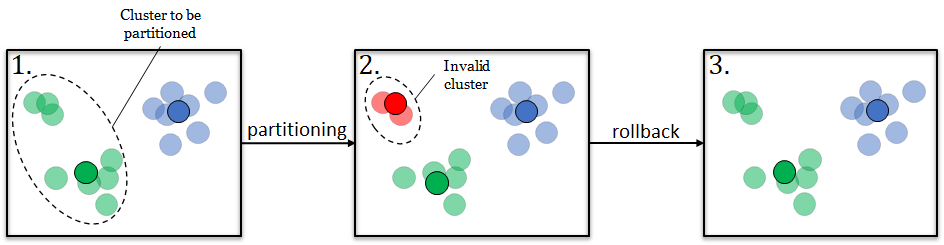}\label{fig:processo-fundir-caso2}}
    
    \caption{Validating clusters according to the constraint of minimum cluster size ($\pi$).}
    \label{fig:processo-fundir}
\end{figure}

Finally, due to the requirement that each cluster has at least $\pi$ data points, a few invalid clusters with a smaller size than $\pi$ could appear during hierarchy creation. A rollback process merges the invalid cluster to its nearest cluster based on the complete linkage distance to solve such a problem. For instance, the merging process happens under two circumstances. First, the partition of a cluster with greater size than $\pi$ may generate two invalid clusters; as shown in Figure~\ref{fig:processo-fundir-caso1}, the partition (\textbf{1.}) of the cluster represented by red colors generates two invalid clusters (\textbf{2.}), then, we merge these clusters (\textbf{3.}) to become a leaf node/cluster. In the second case, the partition produces only one invalid cluster, represented in Figure~\ref{fig:processo-fundir-caso2} (\textbf{2.}) by red-colored data points; the rollback process merges this cluster to the one that generated it. In the figure, non-transparent circles with a black halo encode representative data points.

\subsubsection{ExplorerTree algorithm}
\label{sec:putting-together}

Algorithm~\ref{alg:explorer-tree} summarizes the main steps for creating the ExplorerTree hierarchical structure. The algorithm starts by receiving a set of points ($X$) in the $\mathbb{R}^2$, the sampling selection strategy, the minimum size of the clusters/nodes $\pi$, and the parent node. The algorithm's first step is to select representative data points from the data, as shown in line 3. Then, each non-representative data point is associated with the closest representative data point (line 6) – we also store the parent information for the nodes created by such association (line 7).

\begin{algorithm}[!htb]
\small
\caption{Creating ExplorerTree structure.}
\label{alg:explorer-tree}
\begin{algorithmic}[1]
\Procedure{$\mathbf{ExplorerTree}$}{$X \in \mathbb{R}^2$, Sampling, $\pi$, parent}

    \LineComment{Representative selection process}
    \State points $\leftarrow$ Sampling($X$)
    \item[] 
    
    \LineComment{Clustering: associate each non-representative data point to the closest representative}
    
    \For{ $i \in |\text{points}|$}
        \State node.points $\leftarrow$ $\lbrace x \in X | \parallel x-point\parallel \leq \parallel x-p_j\parallel, i\neq j\rbrace$
        \State node.parent $\leftarrow$ parent
    \EndFor
    \item[] 
    
    \LineComment{Solving problems with invalid clusters, i.e., $|node| < \pi$}
    
    \For{ node$_i \in$ nodes}
    
        \If{ node$_i$.size $< \pi$}
            \State closest\_node $\leftarrow$ closest node$_j$ based on complete linkage distance ($i \neq j$)
            \State closest\_node $\leftarrow$ closest\_node.points $\cup$ node$_i$.points
        \EndIf 
        
    \EndFor
    \item[]
    
    \LineComment{If $|nodes| = 1$, the parent must be a leaf}
    
    \If{ $|nodes| = 1$}
        \State \Return null
    \EndIf 
    \item[]
    
    \LineComment{If $|nodes| \neq 1$, we probably have more levels}
    
    \For{ node $\in$ nodes}
    
        \If{ node.size $\geq 2\times\pi$}
            \State node.children $\leftarrow$ ExplorerTree(node.points, Sampling, $\pi$, node)
        \EndIf 
    
    \EndFor

\State \Return nodes
\EndProcedure
\end{algorithmic}
\end{algorithm}

As commented previously, the partitioning process (representative selection and clustering) could generate invalid clusters with a size lower than $\pi$. Lines 8-12 of the algorithm check if each cluster generated in lines 6 and 7 is invalid. If it is the case, these clusters are merged to the closest cluster according to the complete linkage metric (mean of distances between data points from the two clusters). After this process, if all the nodes were merged, we return null, and the parent node becomes a leaf (lines 14 and 15). 
Finally, if the clusters/nodes generated from the partitioning process produced valid clusters, we create new hierarchical levels if the nodes' size is bigger than $2\times\pi$. It means that we must have room for a least one partition. Thus, we recursively call the same function with $X$ represented by the points in each node and the parent set to the node. The result of such a call is set as the children of the nodes.

\subsection{Visualization design}
\label{sect:exploration-approach}

\begin{figure}[!htb]
    \centering
    \subfloat[MNIST set with $10000$ data instances projected using t-SNE.]{\includegraphics[width=0.45\linewidth]{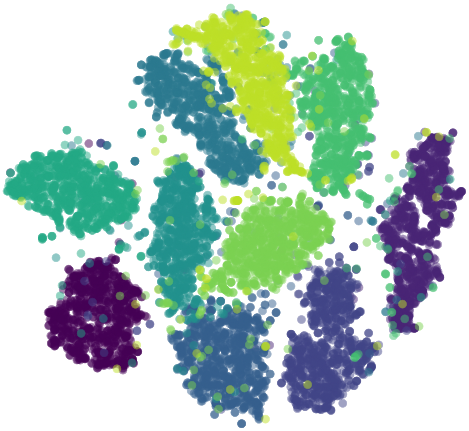}\label{fig:mnist}}    
    \qquad
    \subfloat[SADIRE sampling technique on the MNIST set.]{\includegraphics[width=0.45\linewidth]{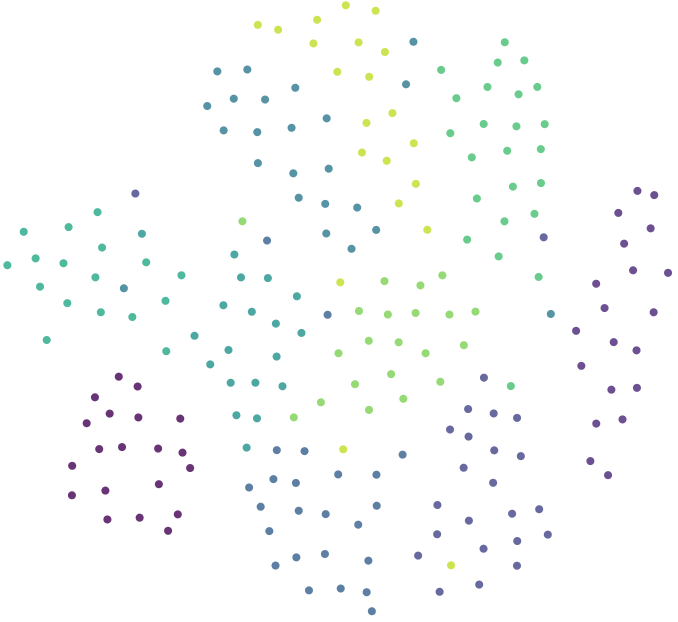}\label{fig:mnist-sadire}}    
    \qquad
    \subfloat[ExplorerTree applied to the embedding of MNIST dataset.]{\includegraphics[width=\linewidth]{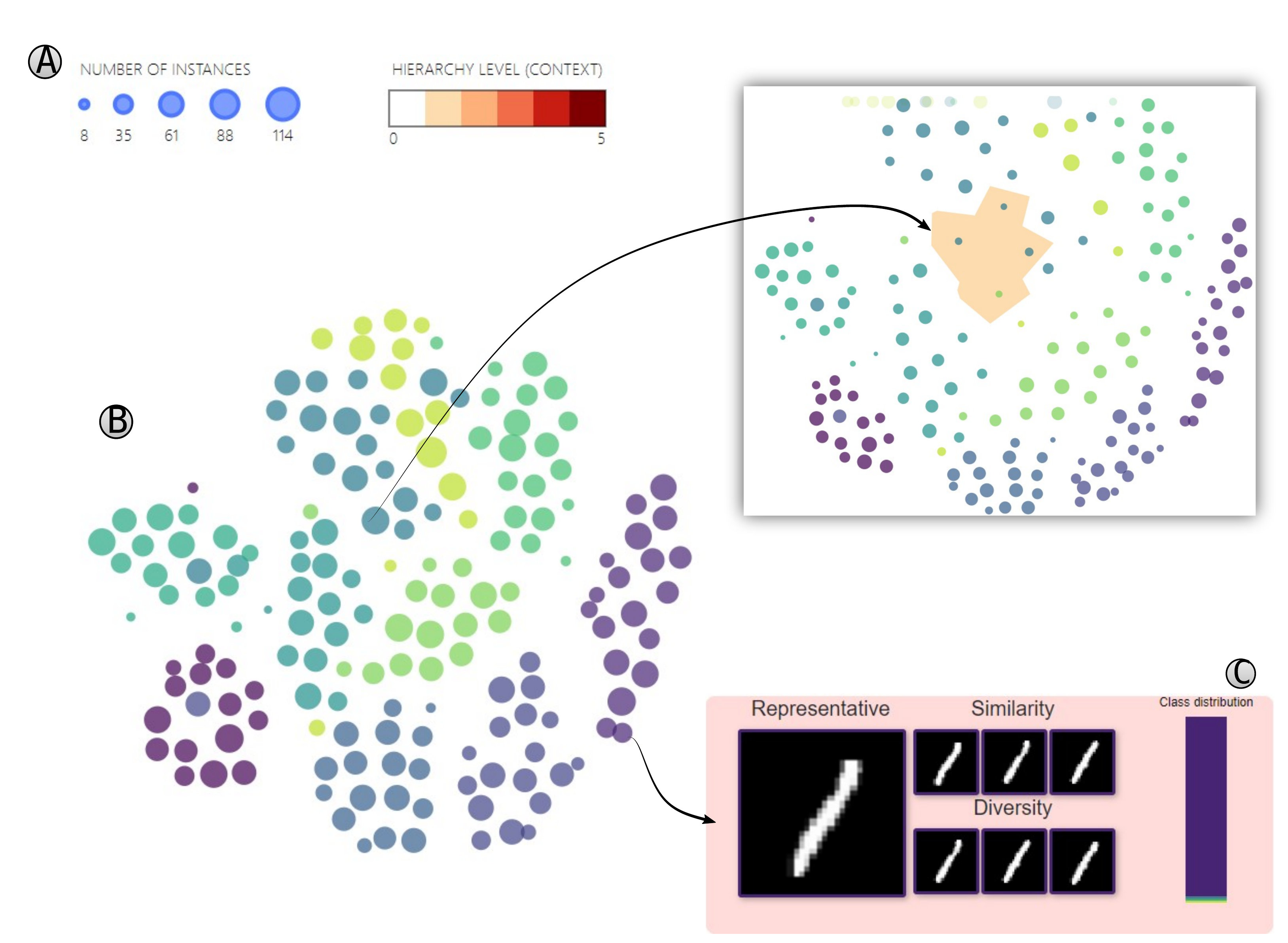}\label{fig:exemplo-sadire}}
    
    \caption{Visual comparison among strategies to encode dimensionality reduction results. Besides interaction, ExplorerTree uses various encoding strategies to add information and help analysts deal with cognitive overload.}
    \label{fig:exemplo}
\end{figure}

In this section, we detail the visualization design of our approach by using a t-SNE~\cite{Maaten_2008} projection of MNIST~\cite{LeCun2010} dataset (see Figure~\ref{fig:exemplo}). After delineating our design choices, we validate the ExplorerTree using the same MNIST dataset by exploring the activation images of convolutional layers of a convolutional neural network trained on it.

As discussed earlier, the traditional approach to visualize dimensionality reduction results (scatter plots) has a few issues regarding visual clutter, interaction mechanisms, and overlapping. While the sampling technique reduces the overplotting caused by the number of visual markers, it loses the density information and makes it difficult to visualize the boundaries among classes (see Figure~\ref{fig:mnist-sadire}). Our approach introduces focus+context interaction to non-hierarchical embeddings, providing a layout with reduced visual clutter and visual guides to explore datasets using representatives. Figure~\ref{fig:exemplo-sadire} shows the first level of the hierarchy built for the embedding of Figure~\ref{fig:mnist}. Notice that we present only the sampled instances (representatives) on the projection plane to reduce visual clutter in \textbf{B}, where each circle area encodes cluster size (see the legend in \textbf{A}). The arrow going from a representative to another embedding depicts the focus request to that representative; the other three representatives' projection and the definition of the focus area highlighted in orange update the embedding, which indicates the hierarchical level. Finally, ExplorerTree shows additional information about the clusters (such as class distribution, similar and diverse data instances) to provide more details about the dataset.

For the following example of interactions, we used the representative selection parameters $k = 15$ and $\alpha = 1$, and the minimum size of clusters ($\pi$) as $5$.

\subsubsection{Requesting Focus}

\begin{figure}[!htb]
\centering
\includegraphics[width=\linewidth]{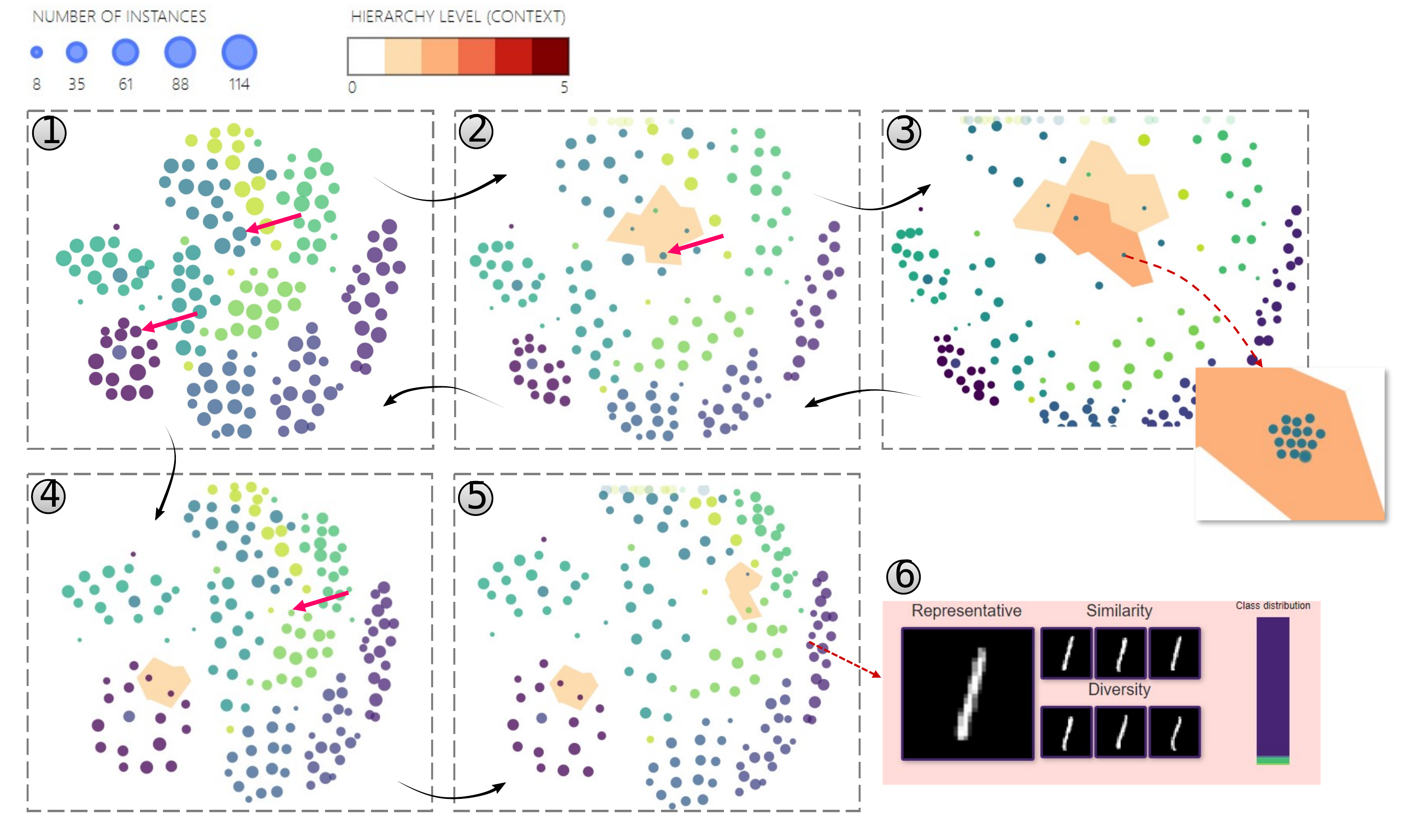}
\caption{Example of interaction with ExplorerTree. In \textbf{(1)} we show the first level of the hierarchy, using circle area to encode cluster size. Layouts \textbf{(2)} and \textbf{(3)} indicate the focus request until reaching a leaf cluster -- see zoomed area in \textbf{(3)}. Layouts \textbf{(4)} and \textbf{(5)} show focus comparison between two clusters. See how color scale of focus level can help users on locating themselves in the hierarchy.}\label{fig:request-focus}
\end{figure}

Users request focus ($\mathscr{F}_+$) through mouse clicks. Figure~\ref{fig:request-focus} illustrates this process when requesting focus from \textbf{1} to \textbf{2}. The focus area is depicted by a different color (an orange color scale) from the context (in white), where color saturation of the polygons delimiting the focus area encodes tree levels – see \textit{HIERARCHY LEVEL (CONTEXT)} and the colored polygons/focus area in Figure~\ref{fig:request-focus}. Notice that the focus area receives more visual space during user interaction, and the context clusters are positioned at the embedding borders to maintain relative distances. The color scale used to encode tree levels and the distortions imposed in the embedding are sufficient to accomplish requirements (\textbf{V1}-\textbf{V4}). Moreover, the request-focus operation accomplishes requirements \textbf{I4}, \textbf{I5a}, and \textbf{I6}.

\subsubsection{Resolving Focus}

Analogously as requesting focus, users interact with the representative instances projected to resolve focus. Notice that resolving focus on arbitrary clusters ($\mathscr{F}_-(\mathscr{D}_{j,i}^{j-1, k})$) results in the cluster parent's projection ($\mathscr{D}_{j-1,k}^{j-2,l}$), as illustrated in Figure~\ref{fig:request-focus}. To resolve focus, users may use any cluster representative like in stage \textbf{(3)}. Users return to the initial hierarchy state and accomplish requirement \textbf{I3} through successive resolving focus operations.

\subsubsection{Comparing Focus}
\label{sec:comparing-focus}

Users use the focus comparison operation ($\mathscr{F}_{c+}(\mathscr{D}_{j,i}^{j-1, k})$) to compare different clusters under focus. Thus, there must be an already focused cluster on the embedding before requesting the focus comparison. Otherwise, such a focus comparison results in a requesting focus operation. Figure~\ref{fig:request-focus} (\textbf{(4)} and \textbf{(5)}) illustrates the process of requesting focus comparison. Firstly, users select an arbitrary cluster represented by the representative class colored in purple and indicated by a pink arrow. After requesting focus on the cluster, the focus area is differentiated from the context, as shown on the embedding \textbf{(4)}. Then, another cluster is chosen (indicated on the embedding \textbf{(5)}) compared to the already selected focus. In this case, the expansion of the space induced by the second focus area is limited, as we will explain later.

\subsubsection{Information about clusters}

Besides using representative instances to reduce overplotting (\textbf{R1}) and provide a visual guide to the most important structures of the embeddings, ExplorerTree guides users through exploratory process using information about the clusters. Figure~\ref{fig:request-focus} (\textbf{6}) presents an example of cluster information, showing the content of the representative (left), the three most similar instances to the representative (top), the similar but diverse data observations (bottom), and the class distribution in the cluster. Notice that the color hue depicts the classes, and bar height encodes the number of data observations in each class. 

Finally, users can use background images for the representative data points (similarly as in Silva et al.'s~\cite{Silva2019} work). Such an approach can be useful when the embedding has well-defined clusters, where users can rapidly see patterns and assess similarity by looking at the representative images.

\begin{figure}[!h]
\centering
\includegraphics[width=0.5\linewidth]{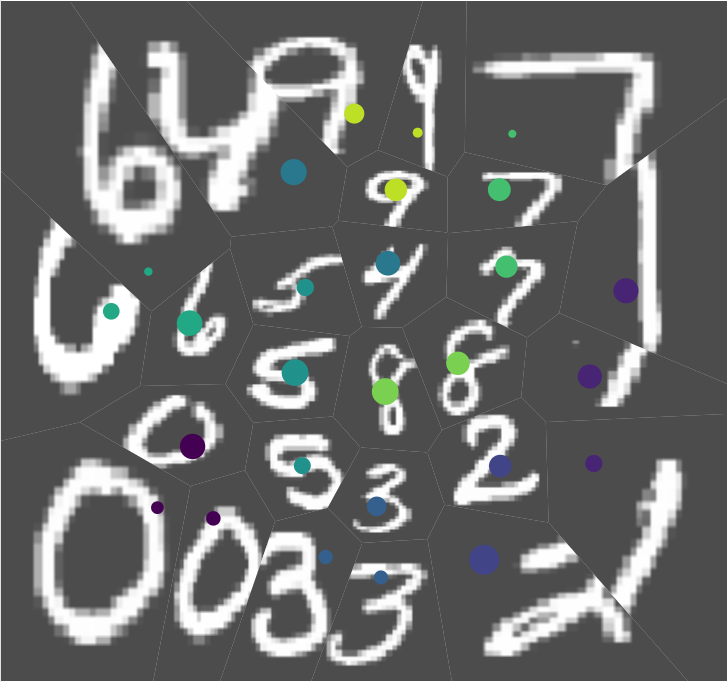}
\caption{Using representative images as background.}\label{fig:background-information}
\end{figure}

\subsubsection{Defining focus and context area}
\label{fig:space-animation}

When requesting focus on a node, the focus area must be defined to distinguish the focus from the context. Besides that, augmenting the focus area also facilitates to distinguish the focus from the context.  Let $x_j^0$ and $y_j^0$ be the coordinates imposed by the embedding technique, such translation is performed using the following equations:

\begin{equation}
\label{eq:translation}
\begin{split}
x_j^i = x_j^{i-1} + (x_j^{i-1} - x')\times f(|\mathscr{D}'|)\times g(\parallel p_j-p'\parallel) \\
y_j^i = y_j^{i-1} + (y_j^{i-1} - y')\times f(|\mathscr{D}'|)\times g(\parallel p_j-p'\parallel)
\end{split}
\end{equation}

\noindent where $x'$ and $y'$ are the coordinates of $p'$, which is the reference/representative point when requesting for a focus; $\mathscr{D}'$ is the cluster that will receive the focus; $p_j$ represent the points in the current embedding and $x_j$ and $y_j$ their coordinates.

Notice in Equation~\ref{eq:translation} the coordinates are interactively updated, i.e., they depend on each request-focus operation held during exploration. Besides, $f$ and $g$ will also play an important role in the positioning of the points. $f$ is a function that linearly maps the clusters' size ($|D'|$) to a value between $0.5$ to $4.0$. $g$ logarithmically maps the euclidean distance ($\parallel p_j-p'\parallel \in [0, \mathscr{M}]$) between the representative point in the request focus operation and the projection points to a value between $0$ and $2.0$ -- $\mathscr{M}$ is the max distance between two points in the embedding. Therefore, $f$ tends to give more space area to clusters with higher density while $g$ tends to reduce the ``pushing out'' effect by giving less importance to clusters as they became far from the focus area. Finally, $g$ also controls the comparing focus process, i.e., $g$ will return $0$ for all the points that are at a distance greater than a limit $\delta$ in pixels (we used $\delta = 100$ in this work). Such control is important since we argue that the focus area must remain constant for comparison.

\section{Use Case - Finding meaningful activations of CNN filters}
\label{sec:use-case}

In this use case, we apply our technique to explore activation images of convolutional filters of a CNN trained on the MNIST dataset, which we already discussed in Section~\ref{sect:exploration-approach}. Here, we aim to inspect discriminative and non-discriminative filters~\cite{Rauber2017,Pezzotti2018} of a convolutional layer. The well-known MNIST~\cite{LeCun2010} dataset has $50k$ training images, $10k$ validation images, and $10k$ test images. We trained a CNN composed by $28\times 28 \times 1$ input image, followed by a convolutional layer with $16$ $3\times 3$ filters, a convolutional layer with $16$ $3\times 3$ filters, a $2\times 2$ max-pooling layer (dropout $0.25$), a fully connected layer, a dense layer with $100$ neurons (dropout $0.5$), and a soft-max output layer with $10$ neurons. The CNN was trained during $500$ epochs, achieving $0.9671$ of test accuracy.

\begin{figure}[!htb]
\centering
\includegraphics[width=\linewidth]{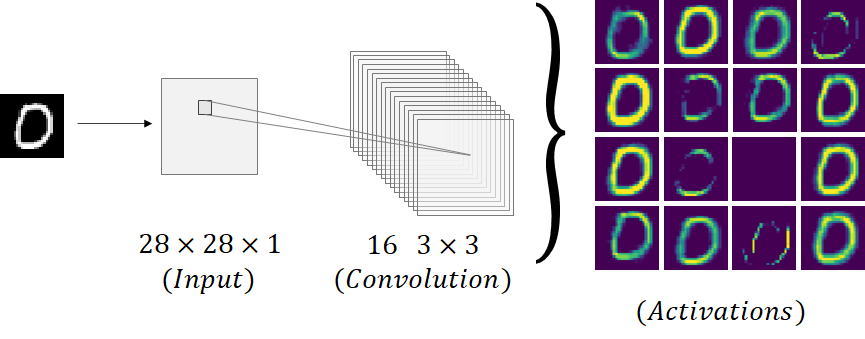}
\caption{For each image in testing set, $16$ activation images were generated.}\label{fig:retrieving-activations}
\end{figure}

We used the test set -- where $600$ images were randomly selected from each class -- to generate each filter's activations from the first convolutional layer. Figure~\ref{fig:retrieving-activations} illustrates the process of generating the activation images for a digit.  To analyze the dataset of activation images through similarity, we used the same model to extract features, for instance, using the fully-connected layer. Thus, the multidimensional matrix has $600\times 10\times 16 = 96000$ rows per $2034$ columns, where $600$ is the number of images retrieved from each class, $10$ is the number of classes, and $16$ is the number of filters.
 
\begin{figure}[h!]
\centering
\includegraphics[width=\linewidth]{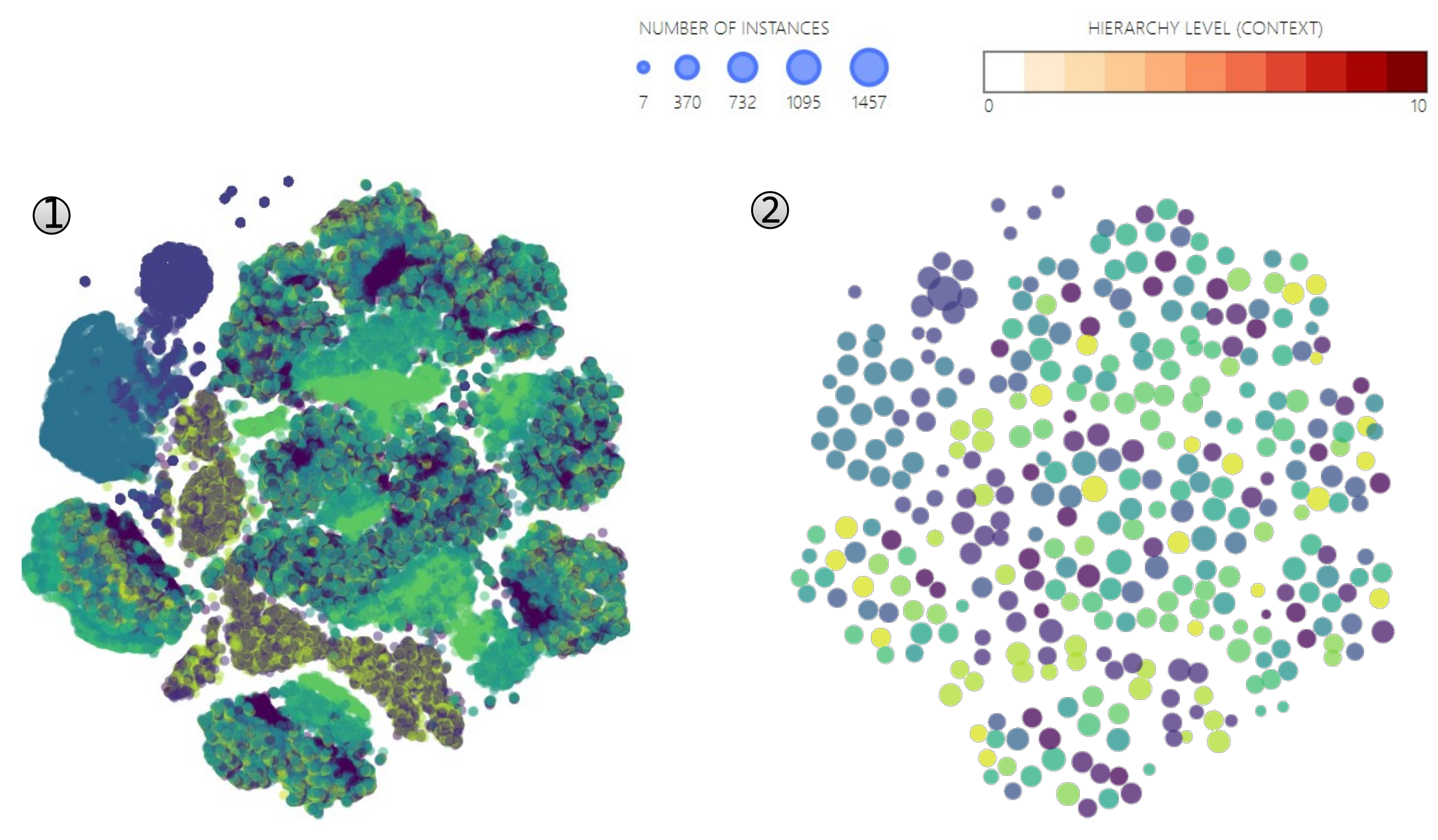}
\caption{Projection of $96000$ of filter activations using colors depict the filters. \textbf{(1)} Projection of each data point. \textbf{(2)} Projection using ExplorerTree.}\label{fig:activations-firstlevel}
\end{figure}

Figure~\ref{fig:activations-firstlevel} shows a scatter plot representation of the filters without sampling next to the first level of ExplorerTree applied to the dataset, for $k = 10$ and $\alpha = 1$. It is worth emphasizing that we are not visualizing the MNIST dataset's images but the activation images generated by the convolutional layer of a CNN. We can see three major structures according to the colors that depict $16$ different filters. Notice how our technique convey most of the traditional projection structures while reducing overplotting imposed by overlapping visual markers.
 
\begin{figure}[!htb]
\centering
\includegraphics[width=\linewidth]{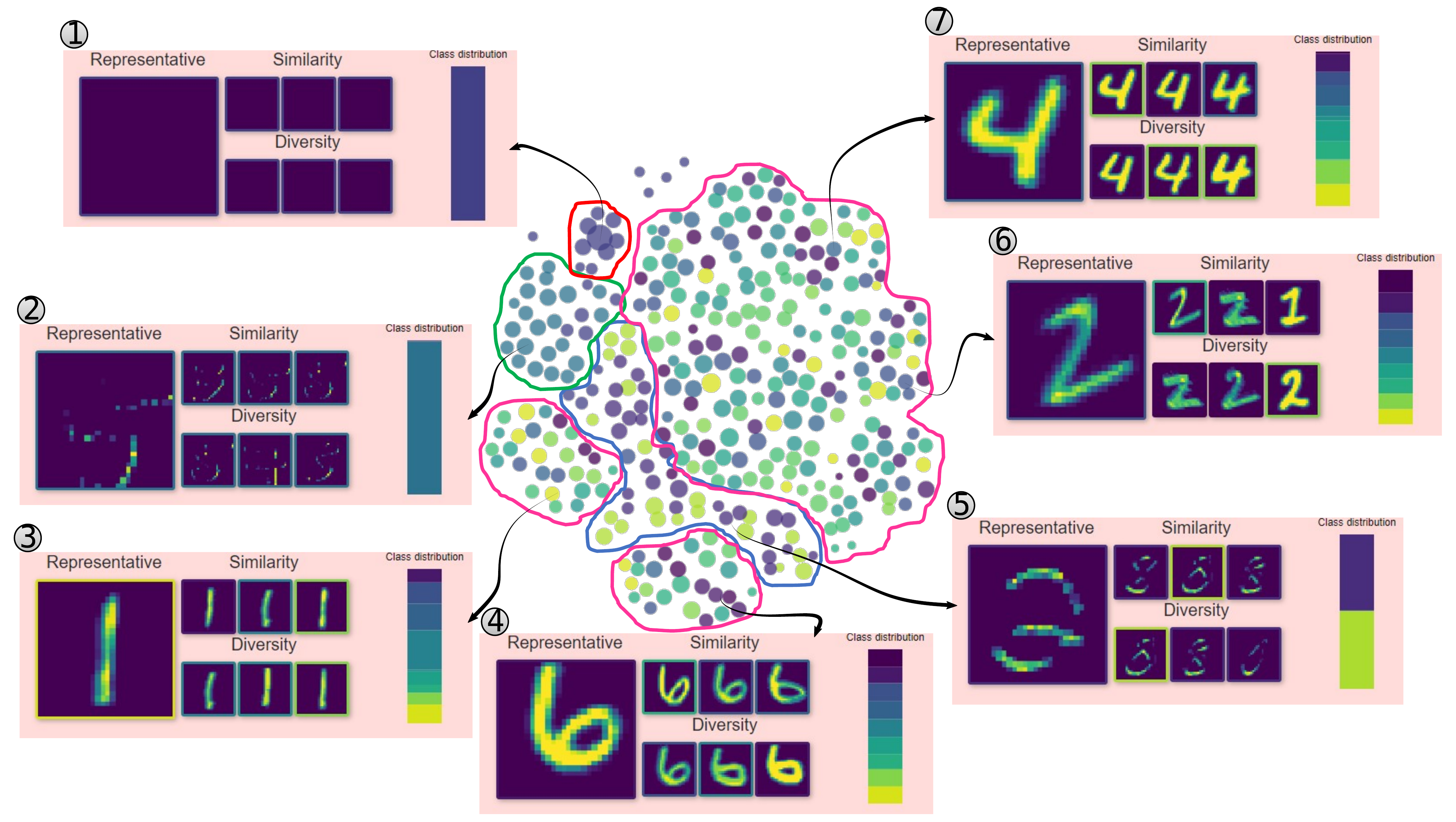}
\caption{Using cluster summaries to inspect the embedding. The summaries emphasize four major structures. Clusters inside the pink boundary represent filters that activate the most. Clusters inside the blue and green boundaries represent filters that activated for some borders. Finally, clusters inside the red boundary did not activate at all.}\label{fig:exploring-tooltips}
\end{figure}

Figure~\ref{fig:exploring-tooltips} illustrates the usefulness of using tool-tips. Users explore the four structures to get more information regarding the clusters so that the exploration task is more comfortable due to the guiding. Notice that we manually separated the main structures with different color boundaries. Such separation emphasizes the filters’ activations, i.e., the structures inside the pink boundaries specify clusters that had more significant discriminative activations -- that contribute to the CNN classification tasks --, the structures inside the blue and green boundaries activate more for the digit borders. Lastly, the clusters inside the red boundary and some outliers closer to them contain filters that did not activate at all. These filters -- colored by purple -- are candidates for removal since they barely contribute to the CNN accuracy. We notice that the two smaller and well-defined clusters are composed of a limited number of filters. While class distribution helps us assess the purity of the clusters, the similar and diverse instances show that the cluster’s instances are similar to each other. In this case, we can see that some filters are not playing a useful role in the classification (see \textbf{(1)} and \textbf{(2)}). Moreover, other filters activate in some borders (see \textbf{(5)}).

\begin{figure}[!htb]
\centering
\includegraphics[width=\linewidth]{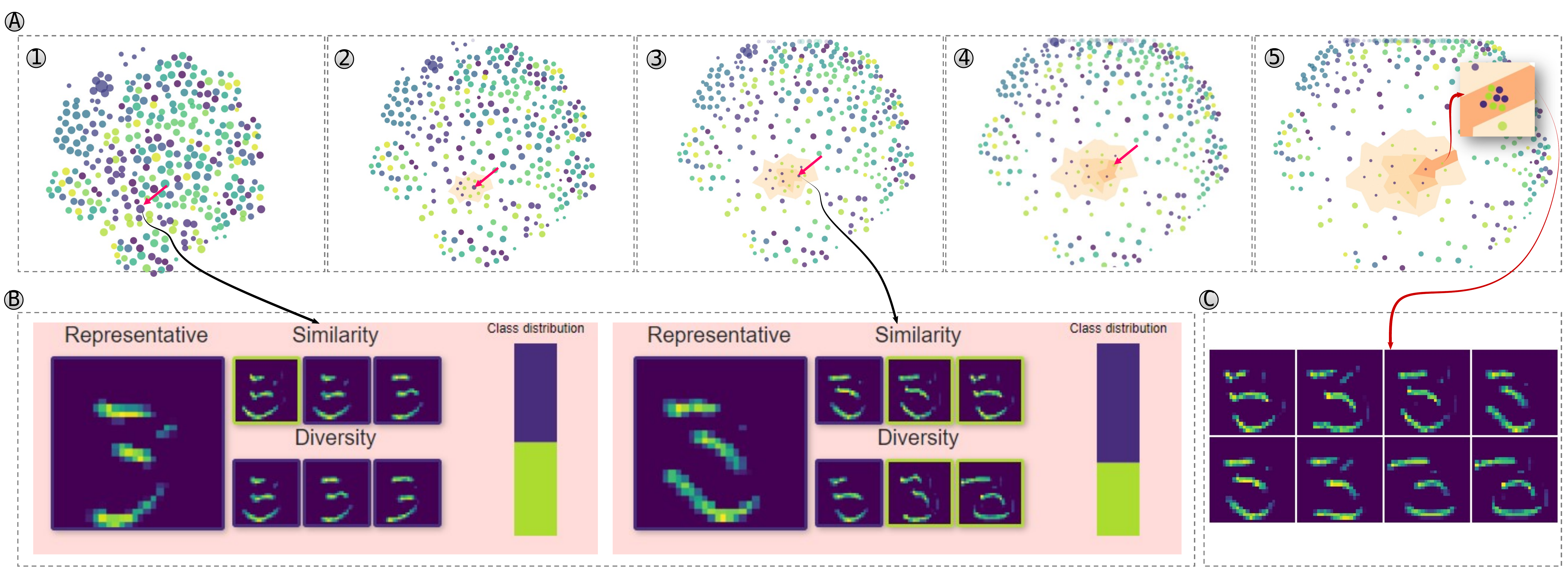}
\caption{Focus requesting operation. \textbf{A)} The embeddings from \textbf{(1)} to \textbf{(5)} show request focus operation  until one reaches a leaf node. Notice that by hovering the leaf cluster, its instances are projected onto the plane. \textbf{B)} The visual summaries help users to assess clusters' content. \textbf{C)} Activation images of leaf cluster.}\label{fig:exploring-activations}
\end{figure}

In Figure~\ref{fig:exploring-activations}A) we select a cluster to inspect and request focus – indicated by a pink arrow. Figure~\ref{fig:exploring-tooltips} shows the cluster positioned inside the blue boundary, which tells that its instances correspond to filters that activate the digits borders. Figure~\ref{fig:exploring-activations}B) (left) show the cluster’s visual summary. The representative instances of the selected cluster \textbf{(1)} are projected in \textbf{(2)}, where we also select a cluster for inspection (see embedding \textbf{(3)}). Notice that as one goes more in-depth in the hierarchy, the number of representative instances to be inspected tends to decrease. Figure~\ref{fig:exploring-activations}B) (right) shows the visual summary of the cluster indicated in the embedding \textbf{(3)}. Comparing both visual summaries, ExplorerTree shows different handwriting traits in the second visual summary and more refined cluster although the main cluster (of embedding \textbf{(1)}) corresponds to digit three activations. Finally, when one reaches a leaf cluster as in \textbf{(5)}, data instances are projected onto the plane without overlapping, where users can inspect their content (in a new window), as shown in Figure~\ref{fig:exploring-activations}C).

\begin{figure}[!htb]
\centering
\includegraphics[width=\linewidth]{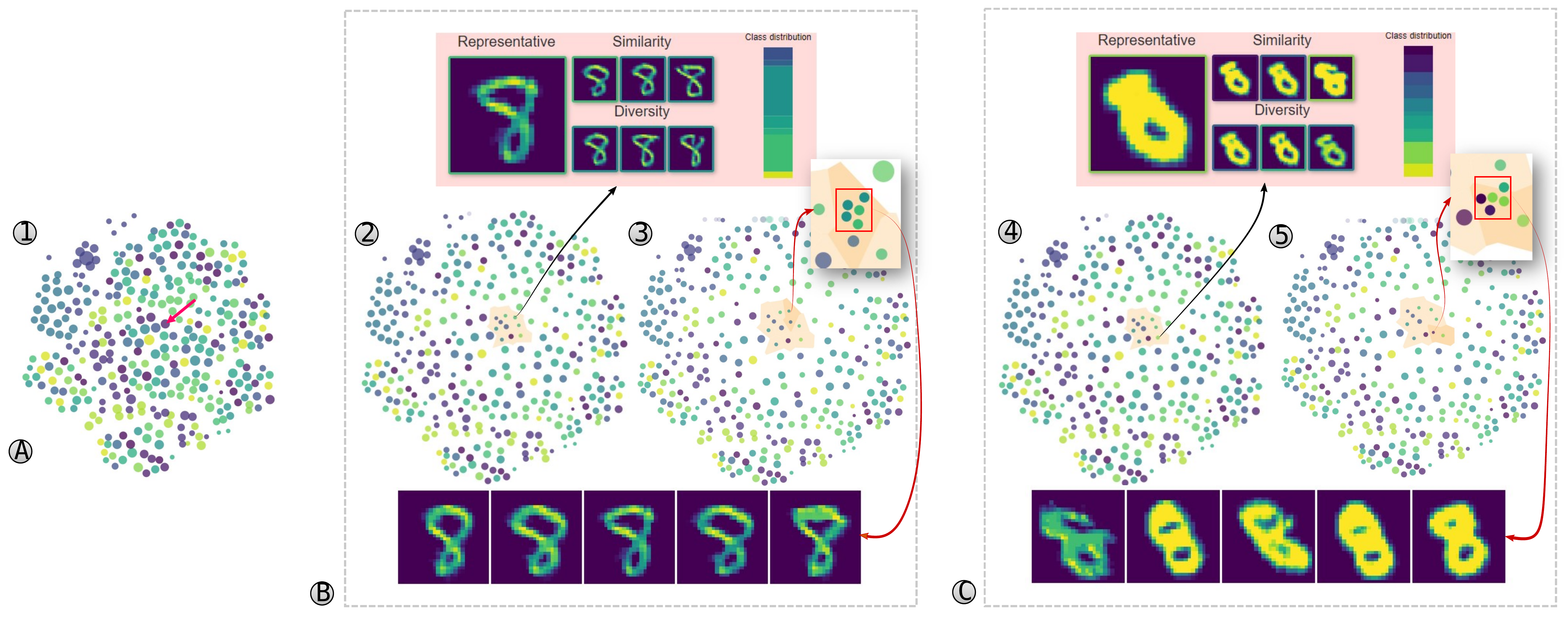}
\caption{Using ExplorerTree to perform intra-cluster analysis. After selecting a cluster of interest \textbf{(A)}, a comparison between two subclusters (\textbf{B.2} and \textbf{C.4}) is performed to investigate cluster homogeneity and similarity. Through investigation of the fine-grained level of two subclusters, activation images show that different convolutional filters result in the same activation pattern.}\label{fig:exploring-activations2}
\end{figure}

After resolving the focus to the hierarchy's first level (i.e., the inverse process of Figure~\ref{fig:exploring-activations}), we illustrate a cluster comparison. In Figure~\ref{fig:exploring-activations2}\textbf{A)} we select a particular cluster to request focus, in which ExplorerTree projects new representative data points onto the plane (\textbf{B}). Using the tool-tip to inspect the cluster information, ExplorerTree shows seven classes inside the cluster, where two of them have the most instances. Moreover, similar and diverse instances show us that the cluster can be populated mainly by instances of digit eight. After requesting more focus results in the projection \textbf{(3)}, ExplorerTree projects two representative data points in the hierarchy's third level. One of these representatives' inspection shows that it corresponds to a leaf node, whose data points are projected onto the plane when recovering the mouse over the representative. These data points, highlighted by a red box in the figure (not part of the visualization), and their correspondent activation images indicated on the bottom of Figure~\ref{fig:exploring-activations2}\textbf{B)} show the cluster homogeneity.

\begin{figure}[!htb]
\centering
\includegraphics[width=\linewidth]{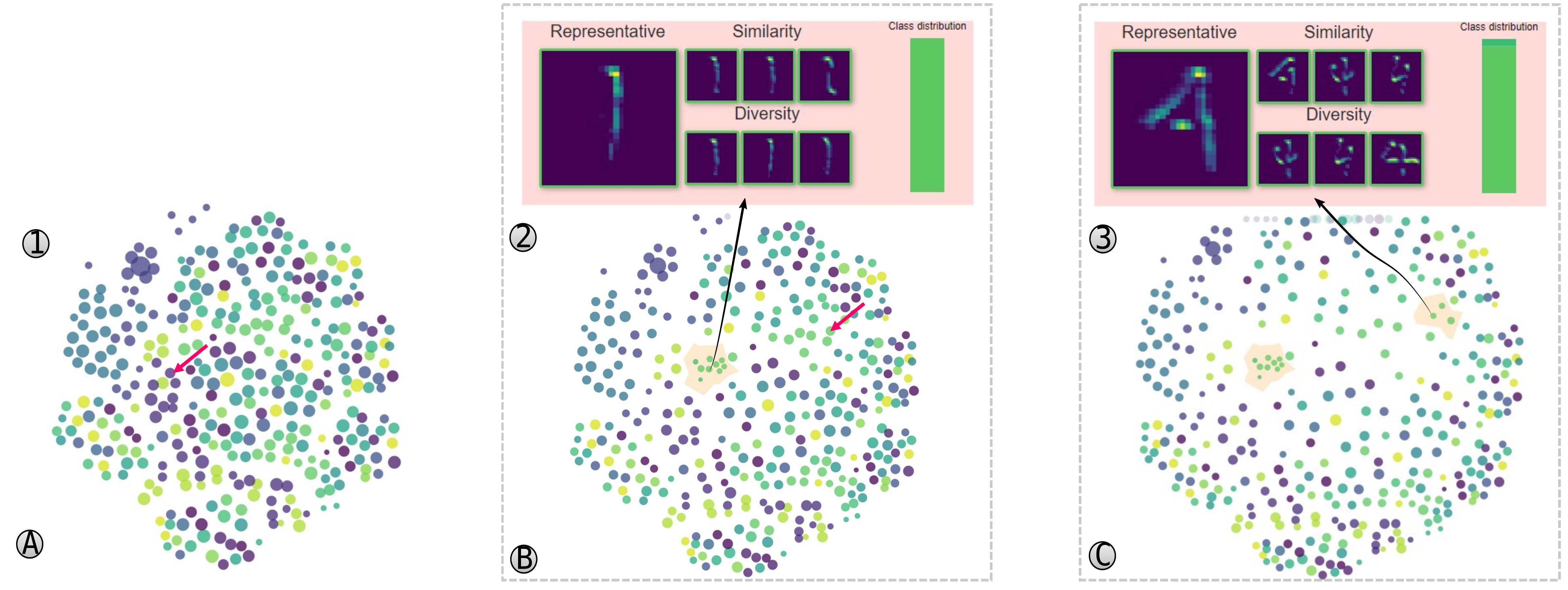}
\caption{Using the operation to compare clusters, ExplorerTree shows how t-SNE can project data samples from the same filter according to the global relationship.}\label{fig:comparing-focus}
\end{figure}

To finish this study case, we use the functionality of comparing clusters in different focus, as shown in Figure~\ref{fig:comparing-focus}. In this case, two clusters that present most data instances generated by the filters encoded as light green were selected – a pink arrow indicates these clusters. After requesting focus on these two clusters and generating the projection of Figure~\ref{fig:comparing-focus}(\textbf{B}) and (\textbf{C}), the summaries (shown on top) of the two inspected data representatives show that such filter is responsible for detecting the digit borders. Notice that, unlike in Höllt et al.'s~\cite{Hollt2019} work, users can freely compare the information of different clusters under focus since the tooltip's information is related to data points, not aggregated structures. By aggregating information using the sampling technique and using visual summaries to show information about the clusters, our technique can quickly inform the structures present in the projection due to the facilitated cluster comparison. For instance, our technique highlights the t-SNE ability to separate the digits even for the same convolutional filter. Such characteristic is impressive since the filter activated for only the digit borders, i.e., the different (even in small amount) traits of digits \textit{one} and \textit{four} were sufficient to place these data points apart.

\section{Discussion and Limitations}
\label{sec:discussion}

Throughout the paper, we demonstrated how our approach could support pattern discovery with reduced cognitive overload. Some of the strengths of our work are: first, our methodology can be easily applied on embeddings generated by any DR technique since it only requires points in $\mathbb{R}^2$; second, the structure provided by our approach allows exploration of a large amount of non-hierarchical projection techniques using the Focus+Context concept; third, the representative selection processes used to define the hierarchy can reduce visual clutter while providing meaningful guides for users – we compensate the lost of information imposed by representative selection through visual summaries of the clusters' information. 

One important thing to keep in mind is that to impose a hierarchical structure on the dataset. We used clustering on the visual space, and a common approach used extensively in the literature for conveying \textit{word clouds}~\cite{Paulovich2012}, images~\cite{Gansner2013}, and so on. One problem with such an approach is the uncertainty in the analysis due to distortions created by the dimensionality reduction process. However, visualizing the result of clustering in the high-dimensional space would increase the cognitive effort to interpret the layout~\cite{NonatoAupetit} due to the generation of fragmented clusters, especially when exploring hierarchical structures. Although further research needs to address the problems to communicate the distortions present in the hierarchy levels, choosing to ease exploration instead of representing high fidelity to clustering seemed the best option for us. 

The sampling selection algorithm~\cite{MarcilioJr2020} employed to maintain the similarity and structure relations imposed by the dimensionality reduction (DR) process, and it accounts for the ability of the DR techniques to uncover the manifolds (if there is any) and to trustfully represent the global and local relations present in the dataset. Although this dependence on the DR technique may seem like a bad feature, it is a good characteristic since it follows the advances in dimensionality reduction research~\cite{McInnes2018, Moon2019}.

\paragraph{\textbf{Requirements}} While our visual scalability requirements were applied only for scatter plot representations of dimensionality reduction results, they can be extrapolated to other scenarios since many visualization techniques suffer from visual disorder and overplotting. For example, graph layouts may suffer from overplotting of edges, and the parallel coordinates are a classic example of how a significant number of data observations can reduce understanding trends and cluster structures. We believe that our requirements for performing analysis in such scenarios can help in design scalable visualization approaches. While the techniques employed to accomplish the requirements may differ from a technique to another -- edge bundling could aggregate trends in parallel coordinates. At the same time, other visual cues could inform about outliers and important patterns. We believe these requirements are universal when designing scalable visualizations.

\paragraph{\textbf{Implementation}} To accommodate other sampling and dimensionality reduction techniques, we developed our tool1 as modular as possible. In this case, different combinations of dimensionality reduction and sampling techniques can create the ExplorerTree as shown in the Supplementary File. Besides, users will have the opportunity to provide new visualization designs for the ExplorerTree hierarchical structure since it was created based on a Java Spring API. The overlap removal, representative selection, and ExplorerTree are different APIs written in Java that can be used to plug in a graphical user interface – we used D3.js~\cite{Bostock2011} to create our visualization design.

\paragraph{\textbf{Limitations}} One limitation of our approach is the dimensionality reduction inability to generate projections that trustfully represent data in high-dimensional spaces. While it can be challenging to assess a projection quality without a quality metric or any other layout enrichment strategy, our technique could hide information about how bad feature spaces influenced the projection result or simply the DR technique's inefficiency. This limitation is related to some visual cues that experts could rely on the projection to assess a layout, which the sampling selection may hide. Although the visual summaries presented here could help gain some information about clusters' organization, we plan to investigate visualization strategies to communicate cluster structures. A second limitation of our approach could be the window space for visualizing the projection, i.e., depending on the density of overlapping, dataset size, and maximum size of each cluster, successive focus requests can deceive user exploration.


\section{Conclusions}
\label{sec:conclusion}

Dimensionality reduction techniques are great tools for analyzing multidimensional datasets. The scatter plots bring benefits according to the rapid assessment of clusters and patterns. However, when the size of the dataset grows, the visual metaphor does not visually scale well.

This work's main contribution is the definition of a hierarchical exploration approach for non-hierarchical embeddings to reduce visual clutter and allow using the Focus+Context exploration concept. As a second contribution, we also propose applying a sampling selection algorithm – developed to preserve class boundaries and general structures of scatter plot representations of dimensionality reduction techniques – on the creation of tree structures for hierarchical exploration of multidimensional embeddings.

\subsection{Future works}
\label{sect:trabalhos-futuros-cap-conclusao}

An interesting strategy is to allow users to define the first hierarchical level to cluster data according to their needs, such as investigating clusters based on gene expression in single-cell analysis applications. Moreover, given that our approach can be easily changed to visualize other types of datasets, \textit{Tags clouds} could be used as a background to provide information about the clusters when exploring document collections.

We also plan to investigate more about the visual scalability requirements we proposed by formulating them in a framework.

\section*{Acknowledgements}

This work was supported by FAPESP (São Paulo Research Foundation),
grants \#2016/11707-6, \#2017/17450-0, \#2018/17881-3, and
\#2018/25755-8.

\bibliographystyle{abbrv}
\bibliography{cas-refs}

\end{document}